\documentclass[11pt]{article}
\usepackage{mons}
\usepackage{graphicx}
\usepackage{times}

\newcommand{\MONS}{{\em MONS\/}}
\def\deg{\hbox{$^\circ$}}

\lefthead{T. R. Bedding \& H. Kjeldsen}
\righthead{\MONS{} Star Trackers}
\begin{document}

\title{The \MONS{} Star Trackers%
%
%\enlargethispage{7ex}
\footnote{To appear in {\em Proceedings of the Third MONS Workshop: Science
Preparation and Target Selection}, edited by T.C.V.S. Teixeira and
T.R.~Bedding (Aarhus: Aarhus Universitet).}
}

\author{Timothy~R.~Bedding}
\affil{School of Physics, University of Sydney 2006, Australia}

\author{Hans Kjeldsen}

\affil{Teoretisk Astrofysik Center, Danmarks Grundforskningsfond,
Aarhus University, DK-8000 Aarhus~C, Denmark}

\begin{abstract}

The \MONS{} satellite will have two Star Trackers to sense the spacecraft
attitude, and we plan to use them as scientific instruments to perform
high-precision photometry of thousands of stars.  We briefly describe the
current plans for the Star Trackers and their expected capabilities.

\end{abstract}

\section{Introduction}

Like many satellites that require precise attitude control, \MONS{} will
use star trackers to sense the spacecraft attitude.  A star tracker is
basically a wide-field CCD camera.  In acquisition mode, images of the sky
are compared with a star catalogue to determine the absolute orientation of
the spacecraft.  Once a target is acquired, the spacecraft attitude is
continuously updated in tracking mode.

There is clear scientific interest in using the photometry from a star
tracker to observe variable stars.  Despite the small aperture, the
advantages of space (long observing periods and no atmospheric
scintillation) make such a camera superior to ground-based telescopes for
some applications.  The only example so far has been the work by
%\citename{BCL2000} (\citeyear{BCL2000} and these Proceedings), who are
Buzasi~et~al. (2000 and these Proceedings), who are
making impressive use of the 52-mm star camera on the {\em WIRE\/}
satellite after the failure of the main instrument.  As far as we know,
\MONS{} is the first mission to be designed from the start to use star
trackers for science.  Of course, we must keep in mind that the primary
role of the star trackers is to sense spacecraft attitude with the
precision required for the main camera.

We describe here the plan for the \MONS{} Star Trackers as it stood at the
time of the Workshop (January 2000).  This plan differs slightly from that
described in the \MONS{} Proposal (Kjeldsen, Bedding \&
Christensen-Dalsgaard 1999), and will undoubtedly change
again before the final design is frozen\footnote{One change is that we now
use the term ``Star Tracker'' rather than ``Star Imager.''}.  There are
several factors driving the changes, including: (i)~performance of trackers
as attitude sensors for the main camera, including redundancy issues;
(ii)~volume, thermal and power constraints from the possible presence of
Ballerina; and (iii)~performance of trackers as secondary science
instruments.

%\newpage
\section{Current plans for the Star Trackers}	\label{sec.plan}

\begin{figure}
\centerline{ \includegraphics[draft=false]{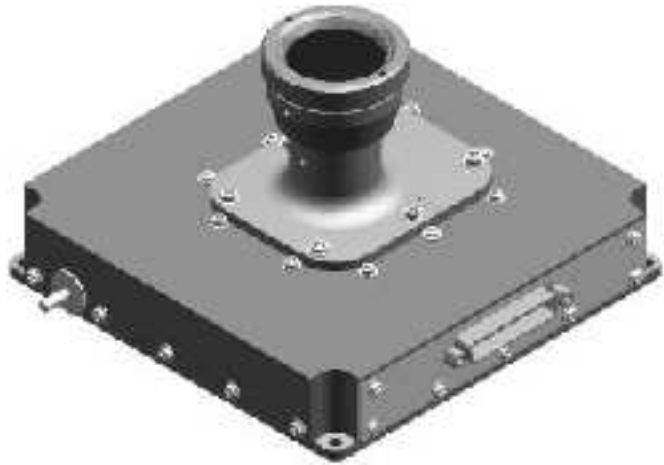} }
\caption[]{\label{fig.chu}  The TERMA Star Tracker.}
%\vspace{-1cm}
\end{figure}

There will be two Star Trackers, pointing in opposite directions.  Both
will consist of a small CCD camera (24\,mm aperture) with a circular field
of view with diameter 22\deg{} (see Fig.~\ref{fig.chu}).  The Bright Star
Tracker (BST) will point forwards, in the same direction as the main
telescope, and will have a broad blue filter (380--420\,nm) to allow it to
observe bright stars without saturating.  The Faint Star Tracker will be
unfiltered, preventing observations of very bright stars but improving the
signal-to-noise at the faint end.

\begin{figure}
\centerline{ \includegraphics[clip=true,width=10cm,bb=95 378 469
662]{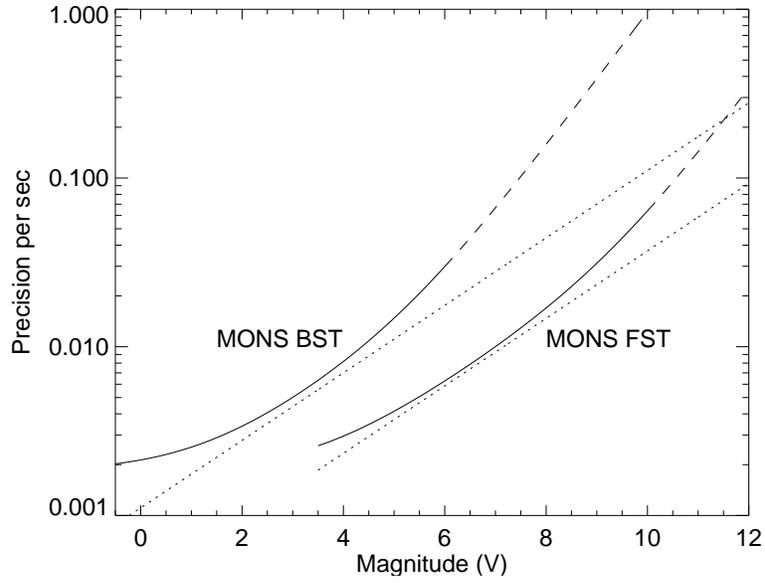} }
\caption[]{\label{fig.a} The expected photometric precision per second for
the two Star Trackers.  The dashed parts of each curve indicate regions
where the noise will probably be too high to give useful data.  The
diagonal dotted lines show the limits set by photon noise.}
\end{figure}

\begin{figure}
\centerline{ \includegraphics[clip=true,width=10cm,bb=95 378 469
662]{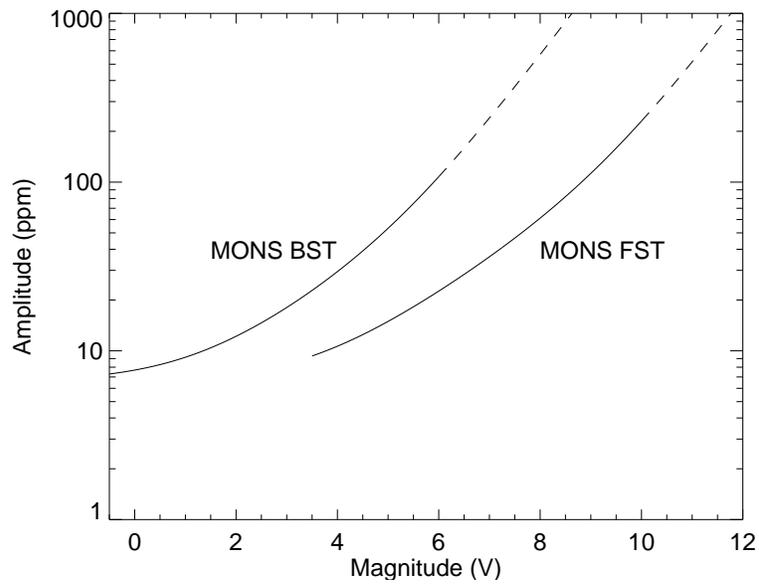} }
\caption[]{\label{fig.b} Expected 3-$\sigma$ sensitivity to oscillations
after 30~days for the two Star Trackers.  }
\end{figure}

\begin{figure}
\centerline{ \includegraphics[clip=true,width=10cm,bb=95 378 469
662]{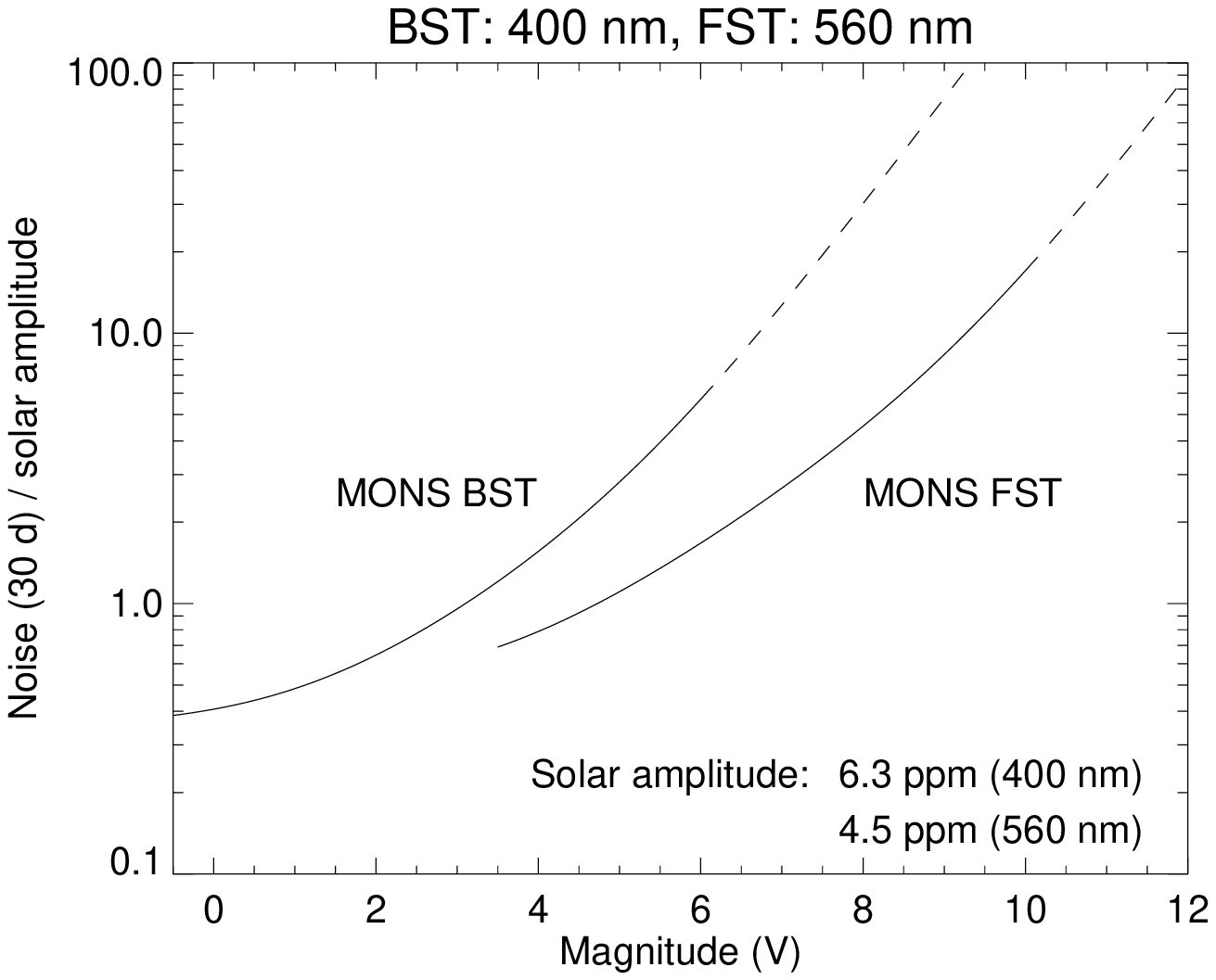} }
\caption[]{\label{fig.c} Expected noise levels after 30~days for the two
Star Trackers, normalised by the solar oscillation signal in the two
passbands.  The BST benefits from the fact that oscillations are stronger
in the blue than the red.}
\end{figure}

Figures~\ref{fig.a}--\ref{fig.c} show the expected photometric performance
of the two Star Trackers.  As a guide, the FST will take several seconds to
reach the same photometric precision for a given star as an individual {\em
Hipparcos\/} measurement.  Of course, each star was only measured by {\em
Hipparcos\/} about 100 times over a few years, whereas \MONS{} will observe
each field almost continuously for about one month.

We intend to use the {\em Hipparcos\/} catalogue at the input catalogue for
\MONS.  There will be about 200 stars in a typical BST field ($V=-0.5$ to
$8.0$), rising to about 600 in the plane of the Milky Way.  For the FST
($V=3.5$ to $12.0$) the corresponding numbers are about 700 and 2000 stars.

The number of stars that will be measured and down-linked will depend both
on crowding and down-link capacity.  As a general rule one can assume that
MONS BST will measure all {\em Hipparcos\/} stars down to $V=7$ in the
non-Milky Way fields, except stars that are too crowded.  For the FST, the
equivalent value is $V=8.5$.  Stars below these magnitude limits will only
be observed if they are of particular scientific interest.

%\newpage
\section{Possible changes to the plan}

Details subject to change include the directions in which the Star Trackers
point and the filters used.  For example, one might consider having the two
Star Trackers pointing in the same direction, one with a red filter and the
other with a blue filter.  This would give colour information, which is of
great scientific value for many types of stars.  The full range of
observable magnitudes could be retained by using bracketed exposure times,
with the CCD being read out in a sequence (e.g., 50\,ms, 200\,ms and
1000\,ms) to sample both bright and faint stars.  The total number of stars
observed in this revised configuration would be similar to that described in
Section~\ref{sec.plan}, but in two colours and with the stars concentrated
in half the number of fields.

\subsection*{Acknowledgements}

We thank the Australian Research Council for financial support, as well as
the Danish National Research Foundation through its establishment of the
Theoretical Astrophysics Center.

\end{document}